# General formalism of non-equilibrium statistical mechanics, path approach.

**Short title: Paths…**

S.G. Abaimov

E-mail: sgabaimov@gmail.com

**Abstract:** In this paper we develop a general formalism of a path approach for non-equilibrium statistical mechanics. Firstly, we consider the classical Gibbs approach for states and find that this formalism is ineffective for non-equilibrium phenomena because it is based on a distribution of probabilities indirectly. Secondly, we develop a path formalism which is directly based on the distribution of probabilities and therefore significantly simplifies the analytical approach. The new formalism requires generalizing the 'static', state quantities of a system, like entropy or free energy potential, to their path analogues. Also we obtain a path balance equation and an equation of equilibrium path. For the distribution of probabilities we obtain a functional dependence similar to the Feynman's path integral formalism of action, only now the role of a Hamiltonian is played by the state entropy. For the production of the state free energy we illustrate a significant dependence on the type of system's connectivity.

## 1. Introduction

A century ago the formalism of equilibrium statistical mechanics has been developed by Boltzmann [1-3] and Gibbs [4, 5] for thermodynamic systems. However, for the case of non-equilibrium phenomena the general formalism is far from having been completely developed. Multiple approaches have been suggested. However, many of them require some additional assumptions or excessive information. For example, Onsager's kinetics requires linearization [6, 7] while the theory of dynamical response, employing the convolution with a response function [8, 9], beside initial and boundary conditions requires the knowledge of the past evolution of a process during some time interval. Our belief is that the general formalism, if would be developed, should inherit the mathematical beauty and simplicity of the Gibbs approach of equilibrium statistical mechanics.

As an example in this paper we consider a system in the canonical ensemble with a thermobath. Other ensembles could be also constructed following similar rules. In section 2 we describe the general characteristics of the system and ensemble. In section 3 we investigate what the distribution of probabilities is for a non-equilibrium system. In section 4 we expand the classical Gibbs approach for non-equilibrium processes and in section 5 demonstrate why this approach becomes ineffective. In section 6 we develop a path approach as the most basic approach directly based on the distribution of probabilities. Also we develop new concepts of path entropy and path free energy potential. For the state free energy in section 6.5 we illustrate why its behavior depends significantly on the type of connectivity of a particular system under consideration. In section 7 we consider the case of random walk of excitations as a simple example.

**2. Non-equilibrium canonical ensemble**

We consider a system in the non-equilibrium canonical ensemble with a thermobath. We will not assume any specific knowledge about the thermobath except that it is in the state of its equilibrium and has many more degrees of freedom than a system and therefore dictates a constant temperature $T$ as a boundary constraint (further on BC) which is independent of any possible system's behavior. The interaction between the system and the thermobath consists only of energy exchange; therefore energy levels of the system and the thermobath are assumed to be identifiable separately.

We assume that the system has discrete energy microlevels $\{E\}$ with energies $E_{\{E\}} \equiv E$. For a system without interactions of degrees of freedom these microlevels are degenerate and packed into macrolevels $[E'] \equiv \bigcup_{\{E\}:E=E'} \{E\}$. For a system with interactions we group closely located microlevels $\{E\}$ into macrolevels $[E]$ as it is usually done in classical equilibrium statistical mechanics. For a macrolevel $[E]$ we assume a general statement that its degeneracy $g_{[E]}$ has an exponential dependence on the number of degrees of freedom $N$ in the system. For simplicity we assume all microprocesses in the system to be Markov of order 1, however we will not require any knowledge about reversibility of these processes.

For both equilibrium and non-equilibrium systems we associate states of a system with its energy levels. So, as microstate $\{E\}$ we will refer to a microlevel $\{E\}$ and as macrostate $[E]$ we will refer to a macrolevel $[E]$. Everywhere further, because we consider only the case of the canonical ensemble, these terms of states and levels will be used intermittently and we will not differentiate them. In the case of other ensembles the macro- and microstates should be associated with the corresponding system bases, for example, energy levels for a given number of particles for the grand canonical ensemble.

For a macrostate $[E']$, following the terminology of our previous publications [10, 11], we introduce extensive quantities like entropy, energy, and Helmholtz energy of this macrostate (for details of this formalism we refer a Reader to these publications). As energy of the macrostate $[E']$ we will refer to the energy $E_{[E']} \equiv \sum_{\{E\}:\{E\}\in[E']} E_{\{E\}} w_{\{E\}} = E'$ of this macrolevel $[E']$. As entropy of the macrostate $[E']$ we will refer to the entropy of a system isolated in this macrostate: $S_{[E']} \equiv -\sum_{\{E\}:\{E\}\in[E']} w_{\{E\}} \ln w_{\{E\}} = \ln g_{[E']}$ where $w_{\{E'\}} = 1/g_{[E']}$ is a probability of a microlevel for a system isolated in the macrolevel $[E']$. As Helmholtz energy of the macrostate $[E']$ we will refer to the Helmholtz energy of a system isolated in this macrostate: $A_{[E']} \equiv \sum_{\{E\}:\{E\}\in[E']} E_{\{E\}} w_{\{E\}} + T \sum_{\{E\}:\{E\}\in[E']} w_{\{E\}} \ln w_{\{E\}} = -T \ln g_{[E']} e^{-E'/T} = -T \ln Z_{[E']}$ where $Z_{[E']} \equiv \sum_{\{E\}:\{E\}\in[E']} e^{-E_{\{E\}}/T} = g_{[E']} e^{-E'/T}$ is the partial partition function of this macrostate [11].

## 3. Distribution of probabilities

To find the distribution of probabilities for the non-equilibrium canonical ensemble we have to recall first how the Boltzmann distribution of energies is obtained in the classical case of the equilibrium canonical ensemble.

The canonical ensemble (equilibrium or not) is defined as a system isolated together with a thermobath. The BC of isolation requires that the total energy, as a sum of the energy of the system $E$ and the energy of the thermobath $E^{thermobath}$, is supported constant: $E^{\Sigma} \equiv E + E^{thermobath} = const$. When the system realizes itself on a microlevel $\{E\}$, this energy $E$ is consumed from the thermobath which otherwise would have the total energy $E^{\Sigma}$: $E^{thermobath} \equiv E^{\Sigma} - E$. Therefore, the thermobath can realize itself on $g^{thermobath}_{[E^{\Sigma}-E]}$ microlevels. For a system in the state of its own equilibrium and in equilibrium with the thermobath all fluctuations from the state of equilibrium are assumed to be an i.i.d. process (the distribution of probabilities of a next state does not depend on a current state of the system). Because all microlevels of the total system Σ are assumed to be equiprobable, and because as a next microlevel the system can choose any microlevel in its spectrum, the probability $w^{BC}_{\{E\}}$ for the system to choose $\{E\}$ as a next microstate is proportional to $g^{thermobath}_{[E^{\Sigma}-E]}$. Defining the entropy of the thermobath macrolevel $[E^{thermobath}]$ as $S_{[E^{thermobath}]} \equiv \ln g^{thermobath}_{[E^{thermobath}]}$ we obtain $w^{BC}_{\{E\}} \propto \exp\left(S_{[E^{\Sigma}-E]}\right)$. Here we used the superscript 'BC' to indicate that this probability is dictated by the equilibrium with the BC $T$. Also we assumed and will assume further on that the Boltzmann constant $k_B$ is chosen to be unity by the appropriate choice of the temperature units.

Because the thermobath has much higher energy than the system we can use a Taylor expansion to obtain

$$w_{\{E\}}^{BC} \propto \exp\left( S_{[E^{\Sigma}]} - \frac{dS_{[E^{thermobath}]}}{dE^{thermobath}}(E^{\Sigma}) \cdot E \right) \text{ or}$$

(1)

$$w_{\{E\}}^{BC} = \frac{1}{Z^{BC}} \exp\left(-E_{\{E\}} / T\right)$$

where $Z^{BC}$ is the partition function of the system in equilibrium with the thermobath $Z^{BC} \equiv \sum_{\{E\}} \exp\left(-E_{\{E\}} / T\right)$.

Now we return to the case of the non-equilibrium system. The difference from the case of the equilibrium system is that the processes are already not i.i.d. but Markov of order 1. Indeed, macromotion in a system cannot attenuate immediately but only through a chain of non-equilibrium states. The excessive heat cannot be transferred immediately to the thermobath but only through a chain of non-equilibrium states. Some processes can be prohibited as irreversible like disappearance of cracks in a solid. In an ideal gas the molecules cannot jump immediately to arbitrary positions but should follow their velocity vectors. Also, to take energy from the thermobath the molecules must be in the vicinity of the thermobath' boundary. Therefore if the system resides on a microlevel $\{E\}$, not all microlevels are possible as a next state $\{E'\}$. Often the possible next microlevels have to be at the adjacent macrolevels. For example, the thermobath could be a radiation field where each molecule of a system can consume a quant with the given frequency to move its system to a next microlevel. Or, contrary, atoms of the thermobath could be two level systems and consume only given amount of energy. Another bright example is a lattice system of two level atoms where the change of microlevels is equivalent to the flip of one site at the boundary between the lattice and the thermobath. This again illustrates the limited connectivity that real systems possess in contrast to the i.i.d. assumption of the equilibrium Gibbs ensemble.

To find the probability for a non-equilibrium system to move from a microlevel $\{E\}$ on a microlevel $\{E'\}$ we follow the derivation above for the equilibrium case. However now the energy of a current state $E$ cannot be returned to the thermobath during only one jump into a new state and the thermobath has a chance to consume only the change of the energy $E - E'$ but not the total $E$. Therefore the probability $w^{BC}_{\{E\}\to\{E'\}}$ of the transition $\{E\} \to \{E'\}$ is proportional now to $g^{thermobath}_{[(E^{\Sigma}-E)-(E'-E)]}$ and we obtain

$$w^{BC}_{\{E\}\to\{E'\}} \equiv w^{BC}(E, E'-E) = \frac{1}{Z^{BC}(\{E\})}\exp(-(E'-E)/T). \tag{2}$$

Here the connectivity of the thermobath has been assumed to be perfect: each thermobath' microlevel can move to any other thermobath' microlevel.

It could seem that we simply transferred the multiplier $\exp(E/T)$ from the partition function $Z^{BC}(\{E\})$, making the last dependent on $E$. However, the situation is more complex than that because now in the partition function we sum only the possible paths $Z^{BC}(\{E\}) \equiv \sum_{\{E\}\to\{E'\}:\{E\}\ is\ given} \exp(-(E'-E)/T)$. In other words, the distribution of probabilities for a Markov process depends on the paths among states but not on the next states. So, we accumulated the complexity of available paths summing over them. Further on for simplicity we will assume that $Z^{BC}(\{E\})$ does not depend on a particular microlevel $\{E\}$ of the macrolevel $[E]$ but only on the energy *E*: $Z^{BC}(\{E\}) = Z^{BC}(E)$. In other words, it is the same for all microlevels $\{E\}$ of the initial macrolevel $[E]$ - all microlevels of a given macrolevel have identical outgoing and incoming connectivity. What should be done if this assumption is not valid will be briefly discussed later at the end of section 6.

## 4. The classical Gibbs approach of states

In this section we apply the classical equilibrium Gibbs approach to the non-equilibrium system. In other words, we identify microstates of the system with its microlevels and macrostates of the system with its macrolevels. The main our goal is to find the probability for the non-equilibrium system to be in a macrostate $[E|_{t=v}]$ at time $t = v$.

Let us assume that at time $t = 0$ the ensemble of systems is in a macrostate $[E|_{t=0}]$ with the equiprobable initial probabilities $w^{BC}_{\{E|_{t=0}\}} = 1/g_{[E|_{t=0}]}$ for the microstates of this macrostate $[E|_{t=0}]$, where $g_{[E|_{t=0}]}$ is the degeneracy of this macrostate. For simplicity we assume that a jump into a next state takes a discrete time interval $dt = 1$. First we need to find the probability that this system will choose a macrostate $[E|_{t=1}]$ as its next state. For each path from one of the initial microstates $\{E|_{t=0}\}$ into one of the next microstates $\{E|_{t=1}\}$ the probability is the same

$$w^{BC}_{\{E|_{t=0}\}\to\{E|_{t=1}\}} \equiv w^{BC}\left(E|_{t=0}, E|_{t=1} - E|_{t=0}\right) = \frac{1}{g_{[E|_0]}} \frac{1}{Z^{BC}(E|_{t=0})} \exp\left(-(E|_{t=1} - E|_{t=0})/T\right). \tag{3}$$

The total probability for the system to move into the macrostate $[E|_{t=1}]$ equals then the number of all possible paths $g_{[E|_0]} g_{\{E|_{t=0}\}\to[E|_{t=1}]}$ leading from a microlevel of the macrolevel $[E|_{t=0}]$ onto the microlevels of the macrolevel $[E|_{t=1}]$ times the probability of each path given by equation (3):

$$W^{BC}_{[E|_{t=1}]} = g_{\{E|_{t=0}\}\to[E|_{t=1}]} w^{BC}\left(E|_{t=0}, E|_{t=1} - E|_{t=0}\right). \tag{4}$$

An important fact here is that we have introduced an important quantity $g_{\{E\}\to[E']}$ characterizing the 'geometric', Markov connectivity between two macrolevels. This quantity depends only on the connections of the Markov process as a model input and contains all complexity of the process memory.

We were lucky above to obtain such a simple form for equation (4). This happened because the initial probabilities of the microstates were equiprobable: $w^{BC}_{\{E|_{t=0}\}} = 1/g_{[E|_{t=0}]}$. But we encounter a difficulty already at the next time-step. To find the probability for the system to be at time $t = 2$ in a macrostate $[E|_{t=2}]$ we need to sum all possible ways leading the system from a microstate $\{E|_{t=0}\}$ into a microstate $\{E|_{t=1}\}$, and then into a microstate $\{E|_{t=2}\}$, each with its own probability:

$$W^{BC}_{[E|_{t=2}]} = \sum_{[E|_{t=1}]} g_{\{E|_{t=0}\}\to[E|_{t=1}]} w^{BC}\left(E|_{t=0}, E|_{t=1} - E|_{t=0}\right) g_{\{E|_{t=1}\}\to[E|_{t=2}]} w^{BC}\left(E|_{t=1}, E|_{t=2} - E|_{t=1}\right). \tag{5}$$

Correspondently, the probability for the system to be at time $t = v$ in a macrostate $[E|_{t=v}]$ includes already $v-1$ sums over all intermediate macrostates:

$$W^{BC}_{[E|_v]} = \sum_{[E|_1]} \dots \sum_{[E|_{v-1}]} g_{\{E|_0\} \to [E|_1]} w^{BC}\left(E|_0, E|_1 - E|_0\right) \cdot \dots \cdot g_{\{E|_{v-1}\} \to [E|_v]} w^{BC}\left(E|_{v-1}, E|_v - E|_{v-1}\right). \tag{6}$$

To find a state at which the ensemble will be on average at the time $t = v$ we have to find the maximum of equation (6) over all possible macrostates $[E|_v]$. It could be possible for the system to find analytically $g_{\{E\} \to [E']}$ as a number of all possible paths connecting two macrolevels. This number is a model input and therefore is usually known. However, to obtain an analytical solution for $t-1$ sums in (6) is generally a cumbersome or impossible task.

## 5. Restrictions of the classical Gibbs approach

Gibbs statistical mechanics identifies microstates of the system with the system's microlevels and macrostates of the system with the system's macrolevels. With this definition in the general case we have to utilize the general Gibbs formula

$$\frac{dw^{BC}_{\{\}}}{dt}(t) = F\left[w^{BC}_{\{\}}(t)\right] \tag{7}$$

that the evolution of the distribution of probabilities of microstates is determined by some functional dependence on the previous history of this evolution. We assumed in equation (7) that the process is a Markov process of order 1 and the evolution depends only on the current state of the system at this time *t*. Being even more general, we would have to include the functional dependence on the total evolution before the time $t$: $\frac{dw_{\{\}}^{BC}}{dt}(t) = F\left[w_{\{\}}^{BC}(t'), t' \le t\right]$. For a system of theoretical mechanics this equation would be $\frac{dw_{\{\}}^{equil}}{dt}(t) = \left[H, w_{\{\}}^{equil}(t)\right]$, where $H$ is a Hamiltonian of this system. For a quantum system we similarly have $i\hbar\frac{d\widehat{w}_{\{\}}^{equil}}{dt}(t) = \left[\widehat{H}, \widehat{w}_{\{\}}^{equil}(t)\right]$.

To obtain the distribution of probabilities for microstates at time $t$ we have to integrate equation (7) over all possible paths among levels. Any microlevel at the time $t$ is a result of the tremendous number of different paths leading to this microlevel. We have to integrate all these combinatorial paths with their own probabilities to obtain the final distribution of probabilities for microlevels at the time $t$. The probabilities of paths depend on all intermediate macrolevels and not only on the final microlevel. Therefore the integration of combinatorial sums becomes cumbersome.

This approach corresponds to classical Gibbs non-equilibrium statistical mechanics, when microstates of a system are identified with the system's microlevels and macrostates of a system are identified with the system's macrolevels. However, we know that the probabilities for a Markov processes are associated not with the states but with the paths among these states (in other words, with transition rates). Therefore it is much easier to find a distribution of probabilities for a macrogroup of paths than for all paths leading to a macrostate. In Gibbs equation (7) the states of a system were chosen as bases, although we see that everything points on the fact that as bases we should choose not the states but the paths [12-20]. In the next section we will see how to develop an approach, different from Gibbs' one, associated not with the system's states but with the system's paths. The benefit of this approach will be that its bases will not include integrals of the system's states of the previous system evolution but, in contrast, will be this evolution itself.

## 6. Path approach

### 6.1. Distribution of probabilities

As an illustration, we consider again the non-equilibrium canonical ensemble. We assume that each system makes $v$ jumps among its states. For simplicity we assume all time intervals $dt$ be the same for all systems in the ensemble. So, the time evolution of our ensemble has $v$ discrete time-steps from $t_0 = 0$ to $t_v = v \cdot dt$. For each time-step $t_i$ the value of the energy at this time-step we denote as $E|_i$. For the total process the total history of the energy is $E(t) \equiv E|_0,\ldots, E|_v$. We assume that the process always starts from the given value of energy $E|_0$, so the quantity $E|_0$ will not be variable in the ensemble.

At each time-step $t_i$ a particular system in the ensemble has its own value of energy $E|_i$ and is on one of microlevels $\{E|_i\}$ corresponding to this energy. For the total process from $t_0 = 0$ to $t_v = v \cdot dt$ we construct all possible chains of microlevels. Each such chain, as a possible sequence of particular microlevels $\{E|_0\}, \{E|_1\},...,\{E|_v\}$, will be referred to as a micropath $\{E|_0\} \to \{E|_1\} \to ... \to \{E|_v\}$ (further on we will abbreviate this notation as $\{E|_0\} \to \{E|_v\}$).

Let's assume that for a micropath $\{E|_0\} \to \{E|_v\}$ the sequence of the levels has the evolution of the energy $E(t) \equiv E|_0, \ldots, E|_v$. Then the probability of this micropath is

$$w^{BC}_{\{E|_0\} \to \{E|_v\}}[E(t)] = \frac{1}{g_{[E|_0]}} \prod_{i=1}^{v} w^{BC}_{\{E|_{i-1}\} \to \{E|_i\}} = \frac{1}{g_{[E|_0]}} \prod_{i=1}^{v} \frac{1}{Z^{BC}(E|_{i-1})} \exp\left(-(E|_i - E|_{i-1})/T\right) \tag{8}$$

where $Z^{BC}(E) \equiv \sum_{\{E\} \to \{E'\}: \{E\} \text{ is given}} \exp\left(-(E' - E)/T\right)$. This probability $w^{BC}_{\{E|_0\} \to \{E|_v\}}$ is dictated by the prescribed BC $T$. This BC is a model input; a thermobath dictates the equilibrium (with this thermobath) distribution of probabilities for different paths but a system actually can realize a non-equilibrium probability distribution $w_{\{E|_0\} \to \{E|_v\}}$ for its paths. Only the equilibrium distribution of probabilities is dictated by the BC. Therefore we used abbreviation 'BC' to emphasize that this probability distribution corresponds to the equilibrium with the BC.

As to a macropath $[E|_0] \to [E|_1] \to \ldots \to [E|_v]$ (further on we will abbreviate this notation as $[E|_0] \to [E|_v]$) we will refer to a subset of all micropaths $\{E|_0\} \to \{E|_v\}$ corresponding to the specified evolution of the energy $E(t)$: $[E|_0] \to [E|_v] = \bigcup_{\{E_0\} \to \{E_v\}: E_i = E|_i} \{E|_0\} \to \{E|_v\}$. The probabilities of these micropaths are given by equation (8) and the number of these micropaths is

$$g_{[E|_0] \to [E|_v]} = g_{[E|_0]} \prod_{i=1}^{v} g_{\{E|_{i-1}\} \to [E|_i]} \,. \tag{9}$$

where $g_{\{E\} \to [E']}$ is again the number of all possible micropaths leading from a microlevel of the macrolevel $[E]$ onto the microlevels of the macrolevel $[E']$. In section 3 we assumed that the partition function $Z^{BC}(\{E\})$ does not depend on the choice of a particular $\{E\}$ among all $\{E\}\in[E]$, i.e. all initial microlevels of the same macrolevel have the same outgoing connectivity to the next macrolevels. In equation (9) we assume that all incoming connections give rise to the same number of outgoing connections. I.e., we have the same assumption of 'equiconnectivity' at work. Therefore the quantity $g_{\{E\}\to[E']}$ is a 'geometric' model input, defining the structure of the Markov process. What should be done if the assumption of equiconnectivity is invalid we will discuss at the end of this section.

The probability for the system to have a macropath $[E|_0]\to[E|_v]$ (to move among macrolevels with the specified energy $E(t)$) is

$$W^{BC}_{[E|_0]\to[E|_v]}[E(t)] = \sum_{\{\}\to\{\}\in[E|_0]\to[E|_v]} w^{BC}_{\{E_0\}\to\{E_v\}}[E(t)] = g_{[E|_0]\to[E|_v]} w^{BC}_{\{E|_0\}\to\{E|_v\}}[E(t)]. \tag{10}$$

We used above the term 'equilibrium' but did not specify what we refer to using this term. The wrong way would be to imagine a system in some detailed balance. We study the non-equilibrium evolution of the system far from the equilibrium state. Using the term 'equilibrium' we refer to the ensemble of paths whose stochastic properties correspond to the prescribed BC *T* (whose stochastic properties are in equilibrium with the prescribed BC *T*). In other words, if the ensemble of systems chooses its paths in accordance with equation (8), we will refer to these paths as being in equilibrium with the prescribed BC *T*. However, we also can consider other ensembles which do not obey the prescribed BC *T* and follow in their evolutions some non-equilibrium distributions of probabilities $w_{\{\}\to\{\}}$ for paths. These ensembles we will refer to as non-equilibrium.

For the equilibrium we will use two different definitions. The BC *T* is assumed to prescribe the equilibrium probability distribution $w^{BC}_{\{\}\to\{\}}$ for all micropaths. Therefore, the equilibrium with this BC could be identified with an ensemble of systems which realizes all micropaths with equilibrium probabilities (8): $w_{\{\}\to\{\}} = w^{BC}_{\{\}\to\{\}}$. In other words, all micropaths are possible but their probabilities are dictated by the BC *T*. The superscript 'BC' will be used for this definition. Then the value of any time-dependent quantity *f*(*t*) in equilibrium with the BC *T* is by definition

$$\langle f(t)\rangle^{BC} \equiv \sum_{\{\}\to\{\}} w^{BC}_{\{\}\to\{\}} f_{\{\}\to\{\}}(t)\,. \tag{11}$$

In contrast, another definition of the equilibrium is the equilibrium (most probable) macropath, *i.e.,* an ensemble that realizes only (that is isolated in) a subset of micropaths corresponding to the most probable macropath. This is the macropath which gives the main contribution to the partition function. In other words, this is the ensemble which follows only that macropath $[E|_0]\to[E|_\nu]$ which corresponds to the maximum of $W^{BC}_{[E|_0]\to[E|_\nu]}[E(t)]$ in the space of all possible functions *E*(*t*). To distinguish this case the superscript '(0)' will be used.

As an example, we may consider the equilibrium time dependence of energy *E*(*t*). As $\langle E(t)\rangle^{BC}$ we refer to the energy evolution averaged over the equilibrium distribution of probabilities

$$\left\langle E(t)\right\rangle^{BC} \equiv \sum_{\{\}\to\{\}} E_{\{\}\to\{\}}(t) w^{BC}_{\{\}\to\{\}} \equiv \tag{12}$$

$$\equiv \begin{cases} E\big|_0, \; t = t_0 \\ \sum\limits_{\{\}\to\{\}} E_{\{E|_i\}} w^{BC}_{\{\}\to\{\}}, \; t = t_i \end{cases} = \begin{cases} E\big|_0, \; t = t_0 \\ \sum\limits_{[E|_0]\to[E|_\nu]} g_{[E|_0]\to[E|_\nu]} E\big|_i \, w^{BC}_{\{E|_0\}\to\{E|_\nu\}}, \; t = t_i \end{cases}.$$

As $E^{(0)}(t)$ we refer to the dependence of the intact parameter corresponding to the most probable macropath:

$$W^{BC}_{[E|_0^{(0)}]\to[E|_\nu^{(0)}]}[E^{(0)}(t)] = \max_{E|_1,\ldots,E|_\nu} W^{BC}_{[E|_0]\to[E|_\nu]}[E(t)]. \tag{13}$$

Of course, in the thermodynamic limit these quantities are equal: $\left\langle E(t)\right\rangle^{BC} \approx E^{(0)}(t)$.

### 6.2. Path entropy

Now we consider a system isolated in a macropath $[E|_0]\to[E|_\nu]$. The number $g_{[E|_0]\to[E|_\nu]}$ of micropaths corresponding to this macropath is given by equation (9) and the probability of any of these micropaths is $w_{\{E|_0\}\to\{E|_\nu\}} = 1/g_{[E|_0]\to[E|_\nu]}$ (because the system is isolated in this macropath). Because the criterion of isolation in a macropath is not in equilibrium with the BC $T$, the probability, so obtained, does not correspond to the equilibrium distribution for paths (8) and we have not used the superscript ‘BC’. The entropy of this macropath is

$$S_{[E|_0]\to[E|_\nu]} \equiv -\sum_{\{\}\to\{\}\in[E|_0]\to[E|_\nu]} w_{\{\}\to\{\}} \ln w_{\{\}\to\{\}} = \ln g_{[E|_0]\to[E|_\nu]}. \tag{14}$$

We should emphasize here that so introduced entropy is the ‘path’ entropy of the distribution of probabilities for the paths [13, 20] and must not be associated with the classical Gibbs entropy associated with the distributions of probabilities for the states (levels). In our notations of section 2 the classical Gibbs entropy would be $S(t) \equiv -\sum_{\{\}(t)} w_{\{\}(t)}(t) \ln w_{\{\}(t)}(t)$ as the average at time $t$ over the probabilities $w_{\{\}(t)}(t)$ of microlevels at this time. Our path entropy $S \equiv -\sum_{\{\}\to\{\}} w_{\{\}\to\{\}} \ln w_{\{\}\to\{\}}$ is associated with the probabilities $w_{\{\}\to\{\}}$ of micropaths of the total process [13, 20] and cannot be attributed to system characteristics at a particular time $t$.

For the equilibrium with the BC $T$ the distribution of probabilities is the equilibrium distribution (8): $w_{\{\}\to\{\}} = w^{BC}_{\{\}\to\{\}}$. Therefore the equilibrium entropy is

$$S^{BC} \equiv -\sum_{\{\}\to\{\}} w^{BC}_{\{\}\to\{\}} \ln w^{BC}_{\{\}\to\{\}} = -\sum_{[E|_0]\to[E|_v]} g_{[E|_0]\to[E|_v]} w^{BC}_{\{E|_0\}\to\{E|_v\}} \ln w^{BC}_{\{E|_0\}\to\{E|_v\}} . \quad (15)$$

The function $W^{BC}_{[E|_0]\to[E|_v]}$, given by equation (10), is a product of $g_{[E|_0]\to[E|_v]}$ and $w^{BC}_{\{E|_0\}\to\{E|_v\}}$. Both these functions contain an exponential dependence on the number of degrees of freedom $N$ ($N$ is infinite in the thermodynamic limit). Therefore the function $W^{BC}_{[E|_0]\to[E|_v]}$ has a very narrow maximum at the most probable, equilibrium macropath $[E|_0^{(0)}]\to[E|_v^{(0)}]$ (the width of this maximum is proportional to $1/\sqrt{N}$). The number of different macropaths $[E|_0]\to[E|_v]$ in the width of this maximum has a power-law dependence on $N$ while the number $g_{[E|_0]\to[E|_v]}$ of micropaths $\{E|_0\}\to\{E|_v\}$ corresponding to each of these macropaths $[E|_0]\to[E|_v]$ has the exponential dependence on $N$. For the normalization of the function $W^{BC}_{[E|_0]\to[E|_v]}$ we obtain

$$1=\sum_{[E|_0]\to[E|_v]} W^{BC}_{[E|_0]\to[E|_v]} \approx_{\ln} W^{BC}_{[E|_0^{(0)}]\to[E|_v^{(0)}]} \equiv g_{[E|_0^{(0)}]\to[E|_v^{(0)}]} w^{BC}_{\{E|_0^{(0)}\}\to\{E|_v^{(0)}\}}[E^{(0)}(t)]\,. \tag{16}$$

where the symbol "$\approx_{\ln}$" means that in the thermodynamic limit $N\to+\infty$ all power-law multipliers are neglected in comparison with the exponential dependence on $N$. Further on, the symbol "$\approx_{\ln}$" will mean the accuracy of an exponential dependence on $N$ neglecting all power-law dependences (logarithmic accuracy). For the logarithm of such equations we will use symbol "$\approx$".

From equation (16) we conclude that

$$g_{[E|_0^{(0)}]\to[E|_v^{(0)}]} \approx_{\ln} 1/w^{BC}_{\{E|_0^{(0)}\}\to\{E|_v^{(0)}\}}[E^{(0)}(t)]\,. \tag{17}$$

In equation (15) for the equilibrium entropy the function $\ln w^{BC}_{\{E|_0\}\to\{E|_v\}}$ has a power-law dependence on $N$ in comparison with the functions $g_{[E|_0]\to[E|_v]}$ and $w^{BC}_{\{E|_0\}\to\{E|_v\}}$ which have the exponential dependences on $N$. Therefore for the equilibrium entropy we obtain

$$S^{BC} \approx -\ln w^{BC}_{\{E|_0^{(0)}\}\to\{E|_v^{(0)}\}}[E^{(0)}(t)] \approx \ln g_{[E|_0^{(0)}]\to[E|_v^{(0)}]} = S_{[E|_0^{(0)}]\to[E|_v^{(0)}]}\,. \tag{18}$$

### 6.3. Boltzmann formalism

In equation (8) for the probabilities of micropaths we see not only the exponential Boltzmann dependence on energies but also the dependence contained in the partition function $Z^{BC}(E) \equiv \sum_{\{E\}\to\{E'\}:\{E\}\text{ is given}} \exp\left(-(E'-E)/T\right) = \sum_{[E']} g_{\{E\}\to[E']} \exp\left(-(E'-E)/T\right)$. To obtain more clear understanding of the behavior of our ensemble we need to find what the dependence $Z^{BC}(E)$ is. If we look at thermodynamic systems in the non-equilibrium canonical ensembles, like heat conductivity problems, we find that usually the differences of energies between the previous and the next non-equilibrium states are minor. The energy comes in or goes out only through the degrees of freedom in immediate contact with the thermobath. The number of these degrees of freedom $N^{(d-1)/d}$ is related to the total number of degrees of freedom $N$ as the surface $L^{d-1}$ of a lattice with the linear size $L$ is related to the volume of this lattice $L^d$, where $d$ is the dimensionality of embedding space. Therefore we see, that when the system moves into a new state it changes only a minor fraction $E^{(d-1)/d}$ of its energy leaving the bulk volume of the energy $E$ intact. Therefore the relative change of energy is $E^{(d-1)/d}/E = E^{-1/d} \propto N^{-1/d}$. This resembles the situation of equilibrium statistical mechanics when the relative fluctuations of energy have an order of $1/\sqrt{N}$. In that case with logarithmic accuracy we could estimate the partition function as its major term. Similarly, with logarithmic accuracy we can estimate $Z^{BC}(E) \equiv \sum_{[E']} g_{\{E\}\to[E']} \exp\left(-(E'-E)/T\right)$ as its major term $Z^{BC}(E) \approx_{\ln} g_{\{E\}\to[]}$ where $g_{\{E\}\to[]}$ is the number of paths leading from the given microlevel $\{E\}$ onto some nearest or the same macrolevel. We can assume that $g_{\{E\}\to[]}$ is proportional to the degeneracy of this macrolevel $g_{[E]}$:

$$w^{BC}_{\{E|_0\}\to\{E|_v\}}[E(t)] = \frac{1}{Z^{BC}} \exp\left( -E\big|_v / T - \left( \sum_{i=1}^{v} S_{[E|_{i-1}]} \right) / T_{\int} \right) \tag{19}$$

where $S_{[E|_{i-1}]}$ is the state (not path) entropy of the macrolevel $[E|_{i-1}]$, defined in section 2: $S_{[E|_{i-1}]} = \ln g_{[E|_{i-1}]}$. From statistical mechanics we know that the degeneracy

$g_{[E|_{i-1}]}$ grows as a power-law of $E|_{i-1}$ with the exponent being proportional to *N*. For example, for the case of an ideal gas we have $g_{[E]} \approx_{\ln} \left\{const(V/N)^{1/3}(E/N)^{1/2}\right\}^N$ where *V* is the volume. Therefore in the case of the canonical ensemble $T=$ const, $V=$ const, $N=$ const for probability (8) of an ideal gas we obtain

$$w^{BC}_{\{E|_0\}\to\{E|_v\}}[E(t)] = \frac{1}{Z^{BC}} \exp\left( -E|_v/T - \left( \sum_{i=1}^{v} \ln E|_{i-1} \right)/T_{\int} \right) \tag{19a}$$

where $Z^{BC} \propto \exp\left(-E|_0/T\right)$ and $T_{\int}$ is some constant determined by the structure of the energy spectrum. It is easy to see that $Z^{BC}$ is the path partition function of the system $Z^{BC} = \sum_{\{\}\to\{\}} \exp\left\{ -E|_v/T - \left( \sum_{i=1}^{v} \ln E|_{i-1} \right)/T_{\int} \right\}$.

The temperature $T_{\int} \propto 1/N$ of the system is complementary to the integral of the logarithm of the energy evolution: $\sum_{i=1}^{v} \ln E|_{i-1} \propto \int_0^{t_v} \ln E(t)dt$ or to the integral of the entropy evolution: $\sum_{i=1}^{v} S_{[E|_{i-1}]} \propto \int_0^{t_v} S_{[]}(t)dt$. Similar results were suggested by other studies [12, 15, 21] but in those studies it was an integral not of the state entropy but of the total probability (10). In our research (together with [22]) we have narrowed the functional dependence to the integral of the evolution of state entropy. In fact, we redeveloped Feynman's formalism of path integrals, only in our case action is proportional not to the integral of state energy but to the integral of state entropy.

The exponent of probability distribution (19a) and the order parameters appearing in it significantly depend on the system and ensemble considered. Further on for simplicity we will follow only the case of dependence (19a) for an ideal gas, however we should remember that equation (19a), in contrast to equation (19), is system dependent and not universal. Firstly, the structure of the Markov process could significantly change our assumption that $Z^{BC}(E) \approx_{\ln} g_{\{E\}\to[]} \propto g_{[E]}$. Secondly, other ensembles may introduce complications. So, if we change the BC from isochoric $V$ = const to isobaric $P$ = const, we obtain

$$w^{BC}_{\{(E,V)|_0\}\to\{(E,V)|_v\}}[E(t),V(t)] = \frac{1}{Z^{BC}} \exp\left( -E|_v / T - \left( \sum_{i=1}^{v} \ln E|_{i-1} \right) / T_\int^E - PV|_v / T - \left( \sum_{i=1}^{v} \ln V|_{i-1} \right) / T_\int^V \right) \quad (19b)$$

where we are again able to couple order parameters and temperatures, keeping things simple. However if we change our ensemble from the canonical $N$ = const to the grand canonical $\mu$ = const, we cannot already separate $E$, $V$, and $N$ and have to introduce a new, complex order parameter, which for an ideal gas will be $N \ln\left( const (V/N)^{1/3} (E/N)^{1/2} \right)$.

Finally, the situation becomes even more complex if we consider finite spectra like of the Ising model. For the Ising model without spin interactions the energy of a system's microlevel is proportional to the magnetization of this microlevel. Therefore for the degeneracy of a macrolevel we obtain $g_{[E]} = N! \{N(1+E/B)/2\}!^{-1} \{N(1-E/B)/2\}!^{-1}$ where $B$ is external magnetic field. Then the probability distribution (19) becomes

$$w^{BC}_{\{E|_0\}\to\{E|_v\}}[E(t)] = \frac{1}{Z^{BC}} \exp\left( -\frac{E|_v}{T} + \frac{1}{T_\int} \sum_{i=1}^{v} \left\{ \frac{1+E|_{i-1}/B}{2} \ln \frac{1+E|_{i-1}/B}{2} + \frac{1-E|_{i-1}/B}{2} \ln \frac{1-E|_{i-1}/B}{2} \right\} \right)$$

(19c)

Therefore we see that dependence (19a) is not universal and the choice of order parameters depends on the system considered. For the purpose of simplicity we consider only the case of the system whose energy spectrum is similar to that of an ideal gas (19a). For other cases all further formulae can be derived in the similar way.

Because during one time-step the energy exchange with the thermobath is limited to the order of $N^{(d-1)/d}$, we can generally expect that $Z^{BC}(E) \approx_{\ln} g_{\{E\}\to[]}$ in (8) (the partition function is determined by the energy of the initial state of exodus) and $g_{\{E\}\to[E']} \approx_{\ln} g_{\{\}\to[E']}$ in (9) (the degeneracy is determined by the final state of destination) where generally $g_{\{E\}\to[]} \approx_{\ln} g_{\{\}\to[E]}$. In this case in (10) many multipliers cancel out leaving

$$W^{BC}_{[E|_0]\to[E|_\nu]}[E(t)] \approx_{\ln} \prod_{i=1}^{\nu} \frac{g_{\{\}\to[E|_i]}}{g_{\{E|_{i-1}\}\to[]}} \exp\left(-(E|_i - E|_{i-1})/T\right) = \frac{g_{\{\}\to[E|_\nu]}}{g_{\{E|_0\}\to[]}} \exp\left(-(E|_\nu - E|_0)/T\right) \propto \tag{10a}$$

$$\propto g_{\{\}\to[E|_\nu]} \exp\left(-E|_\nu / T\right).$$

In other words, in this case the probability of the macropath degenerates to the probability of its final state, as if in the equilibrium canonical ensemble, only instead of the degeneracy of that state its incoming connectivity should be used.

### 6.4. Path free energy action

For the system isolated in a macropath $[E|_0]\to[E|_\nu]$ the probabilities of micropaths are $w_{\{E|_0\}\to\{E|_\nu\}} = 1/g_{[E|_0]\to[E|_\nu]}$ and we define the 'action' of the Helmholtz energy for this macropath as

$$\Phi^A_{[E|_0]\to[E|_\nu]} \equiv E|_\nu / T + \left(\sum_{i=1}^{\nu} \ln E|_{i-1}\right) / T_\int - S_{[E|_0]\to[E|_\nu]} =$$

$$= -\ln\left\{ g_{[E|_0]\to[E|_\nu]} \exp\left( -E_\nu / T - \left(\sum_{i=1}^{\nu} \ln E|_{i-1}\right) / T_\int \right) \right\} = -\ln Z^{BC}_{[E|_0]\to[E|_\nu]}, \tag{20}$$

where $Z^{BC}_{[E|_0]\to[E|_\nu]}$ is the partial path partition function [11] of this macropath: $Z^{BC}_{[E|_0]\to[E|_\nu]} = \sum_{\{E|_0\}\to\{E|_\nu\}\in[E|_0]\to[E|_\nu]} \exp\left\{ -E|_\nu / T - \left(\sum_{i=1}^{\nu} \ln E|_{i-1}\right) / T_\int \right\} = Z^{BC} W^{BC}_{[E|_0]\to[E|_\nu]}$. Therefore for the action of the Helmholtz energy we obtain $\Phi^A_{[E|_0]\to[E|_\nu]} = -\ln\left(Z^{BC} W^{BC}_{[E|_0]\to[E|_\nu]}\right)$.

As to the 'action' of the Helmholtz energy we have referred to the quantity $\Phi^A_{[E|_0]\to[E|_\nu]} \equiv E|_\nu / T + \left(\sum_{i=1}^{\nu} \ln E|_{i-1}\right) / T_\int - S_{[E|_0]\to[E|_\nu]}$. A careful Reader notices that if we define the Helmholtz energy as $A_{[E|_0]\to[E|_\nu]} \equiv E|_\nu + T\left(\sum_{i=1}^{\nu} \ln E|_{i-1}\right) / T_\int - TS_{[E|_0]\to[E|_\nu]}$, its action is just the free energy divided by the temperature $\Phi^A_{[E|_0]\to[E|_\nu]} \equiv A_{[E|_0]\to[E|_\nu]} / T$. Why we decided to utilize the action instead of the habitual Helmholtz energy itself we discuss in Appendix A.

The true free energy potential that should be maximized for paths is the probability of these paths $W^{BC}_{[E|_0]\to[E|_\nu]}$, given by equation (10). However, $T_\int$, $T$ and $Z^{BC}$ are

positive constants and the logarithmic function is the monotonically increasing dependence. Therefore we see that the Helmholtz energy or its action $\Phi^{A}_{[E|_0]\to[E|_v]} = -\ln\left(Z^{BC} W^{BC}_{[E|_0]\to[E|_v]}\right)$ can play the role of the free energy potential that should be minimized.

In Gibbs equilibrium statistical mechanics an equilibrium state is found as a minimum of a free energy potential. For the equilibrium microcanonical ensemble the free energy potential is the negative entropy $-S \equiv < \ln w_{\{\}} > \equiv \sum_{\{\}} w_{\{\}} \ln w_{\{\}}$; for the equilibrium canonical ensemble the free energy potential is the Helmholtz energy $A \equiv < E_{\{\}} > + T < \ln w_{\{\}} > \equiv \sum_{\{\}} w_{\{\}} [E_{\{\}} + T \ln w_{\{\}}]$ (for the canonical ensemble the minimization of the Helmholtz free energy is sometimes referred to as a maximization of the entropy. We discuss the difference in Appendix B). For the case of a general equilibrium ensemble in Gibbs equilibrium statistical mechanics the principle of the minimization of the free energy potential always works because this potential is always proportional to the minus logarithm of the probability distribution with the external boundary constraints as constants of proportionality [for detailed discussion see 10]. We see that the same principle is valid and for the case of non-equilibrium statistical mechanics, only now we have to construct the free energy potential not for states but for paths. So, for the path non-equilibrium microcanonical ensemble the negative path entropy

$$-S \equiv < \ln w_{\{\}\to\{\}} > \equiv \sum_{\{\}\to\{\}} w_{\{\}\to\{\}} \ln w_{\{\}\to\{\}} \tag{21}$$

plays the role of the free energy potential. This ensemble corresponds to the situation when all possible paths are equiprobable.

For the path non-equilibrium canonical ensemble the role of the free energy potential is played by the action of the Helmholtz energy

$$\Phi^{A}(t) \equiv < E(t) > /T + < \int_0^t dt' \cdot \ln E(t') > /T_{\int} + < \ln w_{\{\}\to\{\}} > \equiv \tag{22}$$

$$\equiv \sum_{\{\}\to\{\}} w_{\{\}\to\{\}} \left[ E\big|_{v} /T + \left( \sum_{i=1}^{v} \ln E\big|_{i-1} \right) /T_{\int} + \ln w_{\{\}\to\{\}} \right].$$

So, we see that our non-equilibrium system has two parameters similar to a Hamiltonian: one is the value of the energy at the final microlevel and another is the integral of the logarithm of energy. These two parameters could be called 'path' Hamiltonians or path order parameters.

For the equilibrium macropath $[E|_0^{(0)}] \to [E|_v^{(0)}]$ the action of the Helmholtz energy is $\Phi^{A}_{[E|_0^{(0)}]\to[E|_0^{(0)}]} \approx -\ln Z^{BC}$ which coincides with the equilibrium Helmholtz energy obtained by the averaging of the ensemble in equilibrium with the BC $T$

$$\Phi^{A,BC} \equiv \sum_{\{\}\to\{\}} \left\{ E\big|_{v} /T + \left( \sum_{i=1}^{v} \ln E\big|_{i-1} \right) /T_{\int} \right\} w^{BC}_{\{\}\to\{\}} - S^{BC} =$$

$$= -\sum_{\{\}\to\{\}} \left\{ -\ln w^{BC}_{\{\}\to\{\}} - E\big|_{v} /T - \left( \sum_{i=1}^{v} \ln E\big|_{i-1} \right) /T_{\int} \right\} w^{BC}_{\{\}\to\{\}} = -\ln Z^{BC}. \tag{23}$$

### 6.5. Production of state free energy

For the equilibrium canonical ensemble the sharp minimum of a free energy potential is provided by the competition of two exponential dependences: the Boltzmann probability, decreasing with the increase of energy, and the degeneracy of macrolevels, increasing with the increase of energy. Similar situation we see for the path action (22): the Boltzmann distribution of integrals of state entropies competes with the degeneracy of paths. Therefore at the equilibrium path the path action has a very narrow minimum, corresponding to the very narrow maximum of probability (10). This situation is schematically illustrated in Fig. 1a. This minimum is of the action functional in the space of all possible paths. However, in statistical mechanics we used to describe a system by its state's characteristics, not path's characteristics. For example, many studies have been devoted to the investigation of the state free energy production. (In literature the term 'entropy production' is more common assuming that the state entropy is the free energy potential for a system and a thermobath, isolated together. We, for the reasons illustrated above and in Appendix B, prefer to utilize the 'free energy potential' terminology.) To investigate the behavior of the free energy of states we should return from path characteristics to characteristics of macrostates, previously defined in section 2. For the Helmholtz energy of a macrostate $A_{[E']} \equiv \sum_{\{E\}:\{E\}\in[E']} E_{\{E\}} w_{\{E\}} + T \sum_{\{E\}:\{E\}\in[E']} w_{\{E\}} \ln w_{\{E\}} = -T \ln g_{[E']} e^{-E'/T} = -T \ln Z_{[E']}$ we expect to have the narrow minimum at the equilibrium state, where the equilibrium path ends. For the same reason we can expect that during the path evolution the state Helmholtz energy can only decrease, as it is illustrated in Fig. 1b. However, we do not see how the supposed decrease of the state potential or its action would directly follow from the choice of equilibrium path as a minimum of (20). Indeed, the minimum of (20) does not give us directly that $dA_{[E]}/dt \le 0$ or $d\Phi^{A}{}_{[E]}/dt \le 0$. To understand why, we need to consider some illustrative models.

Firstly, we imagine a system with perfect connectivity where each microlevel can move onto any other microlevel. In other words, all path are allowed and nothing prevents the system to choose the equilibrium macrostate already at the first time-step. This situation corresponds to equilibrium statistical mechanics, when a system immediately jumps into the equilibrium state without any intermediate steps. The state free energy already at the first time-step decreases to its minimal value, as it is illustrated in Fig. 1c. Its production rate $dA_{[E]}/dt$ is a negative delta-function, illustrated in Fig. 1d.

For the second model process we assume that the connectivity of microlevels is such that the equilibrium path has the same shift of energy at each time-step. For example, we can imagine a boundary of a thermobath such that at each time-step it emits or consumes only one quant of energy, and each time this quant has a prescribed constant value of energy. In this case the equilibrium path will switch macrolevels with this quant of energy as a constant shift at each time-step. This situation is illustrated in Fig. 1e. Due to the 'bell' shape of the minimum of the state free energy, the production of the state free energy is initially negative, and its absolute value initially increases but then decreases to zero, as illustrated in Fig. 1f. These effects of the initial increase of negative production followed by the steady regime and then followed by the decrease of negative production are all the effect of the limited connectivity. For a macrostate far from the equilibrium state the product $g_{[E]}e^{-E/T}$ is almost zero, therefore the state Helmholtz energy of this macrostate $A_{[E]} = -T \ln g_{[E]}e^{-E/T}$ is zero for many initial time-steps. So, the absolute value of its production rate is initially small but increases. The maximum steady production rate corresponds to the steepest slope of the minimum of the state free energy. And then, at the bottom of this minimum, the production decreases to zero along the equilibrium path.

The constant value of energy shift for each time-step above could be substituted by other connectivity conditions. For example, for an ideal gas its molecules after collisions with the thermobath boundary gain or loose the amount of energy which on average as significant as the system is far from the equilibrium. Indeed, a slow-moving molecule of a cold gas would take more energy from the hot thermobath wall than a molecule of a gas which is as hot as the thermobath. Therefore as a third possible alternative we could consider a situation where energy shifts are larger far from the equilibrium state and smaller in its vicinity. This model would give the behavior analogous to the previous case.

Summarizing the case of the limited connectivity, some paths are prohibited; therefore the system approaches the equilibrium state only after some number of steps. These steps approach the minimum of the state free energy potential. This corresponds to the traditional point of view that the most-probable process in the system always decreases the state free energy potential. The crucial fact here is that the degeneracy of paths between two close macrolevels has an order of the degeneracy of these macrolevels. Only this provides the tendency for a system to decrease its state free energy potential. If this would not be true, the behavior of a system could be anomalous.

For example, as the last possible alternative of connectivity, we could imagine a system whose degeneracy of macrolevels $g_{[E]}$ would increase with the increase of energy of these macrolevels, but whose connectivity of macrolevels would decrease with the increase of energy. In other words, the higher the energy, the higher the degeneracy of the spectrum, but the lower the number of connections of each microlevel to other microlevels. The state of global equilibrium $d(g_{[E]}e^{-E/T})/dE\Big|_0 = 0$ in such a system would have some finite, non-zero energy $E_0$, but this state would be unstable. Any perturbation of this state would cause a transient process, leading the system to zero energy and increasing the state free energy potential. In other words, the minimum of the state free energy potential in this case is unstable, and the system tends, contrary to classical expectations, to increase its free energy. As an example, we can imagine a system which can consume energy from a thermobath only by collisions of its molecules with the boundary but which can emit energy back to the thermobath by photon emission. The mechanism of energy consumption only by molecules in the vicinity of the boundary cannot prevail the connectivity of energy emission when each molecule can emit a quant. Therefore, in spite that the thermobath dictates the higher temperature to the gas, the last will be much colder due to inverse connectivity.

Therefore, in non-equilibrium statistical mechanics the behavior of the state free energy significantly depends on the connectivity type of a particular system under consideration and it is irrelevant to discuss the behavior of the state free energy without a reference to a particular case of connectivity. A system can increase or decrease its free energy production rate during the same process, and this production can become even positive. This has happened because we returned from the path' quantities to the state's quantities. It seems that each time we try to return to simpler for us state point of view, non-equilibrium statistical mechanics makes such returns cumbersome and tangled. When we consider only the path formalism, everything immediately becomes straightforward. Indeed, any system has a strong tendency to minimize its path free energy, but for the state free energy the answer is already not so straightforward.

### 6.6. Balance equation

At the point of the maximum of $W^{BC}_{[E|_0]\to[E|_\nu]}$ (which corresponds to the equilibrium macropath $[E|^{(0)}_0]\to[E|^{(0)}_\nu]$) we have

$$\frac{\partial W^{BC}_{[E|_0]\to[E|_\nu]}}{\partial E|_i}[E^{(0)}(t)]=0 \text{ or } \frac{\partial \ln W^{BC}_{[E|_0]\to[E|_\nu]}}{\partial E|_i}[E^{(0)}(t)]=0 \qquad (24)$$

(where $i\geq 1$ as we assume $E|_0$ not to be variable). For $\ln W^{BC}_{[E|_0]\to[E|_\nu]}=\ln g_{[E|_0]\to[E|_\nu]}+A^{BC}-E|_\nu/T-\left(\sum_{i=1}^{\nu}\ln E|_{i-1}\right)/T_\int$ we can write that

$$\frac{1}{E|^{(0)}_i T_\int}=\left.\frac{\partial \ln g_{[E|_0]\to[E|_\nu]}}{\partial E|_i}\right|_{E^{(0)}(t)}=\frac{\partial \ln g_{[E|^{(0)}_0]\to[E|^{(0)}_\nu]}}{\partial E|^{(0)}_i}, i=1,...,\nu-1$$

and (25)

$$\frac{1}{T}=\frac{\partial\ln g_{[E|_0]\to[E|_\nu]}}{\partial E|_\nu}\Bigg|_{E^{(0)}(t)}=\frac{\partial\ln g_{[E|_0^{(0)}]\to[E|_\nu^{(0)}]}}{\partial E|_\nu^{(0)}}$$

at the equilibrium macropath. These equations could be used as the definitions of the temperatures. As both the entropy of a macropath $S_{[E|_0]\to[E|_\nu]}=\ln g_{[E|_0]\to[E|_\nu]}$ and the path equilibrium entropy $S^{BC}\approx S_{[E|_0^{(0)}]\to[E|_\nu^{(0)}]}$ have the same functional dependence on $E(t)$ and $E^{(0)}(t)$ respectively, we obtain

$$\frac{1}{E|_i^{(0)}T_\int}=\frac{\partial S_{[E|_0]\to[E|_\nu]}}{\partial E|_i}\Bigg|_{E^{(0)}(t)}\approx\frac{\partial S^{BC}}{\partial E|_i^{(0)}},i=1,...,\nu-1 \text{ and } \frac{1}{T}=\frac{\partial S_{[E|_0]\to[E|_\nu]}}{\partial E|_\nu}\Bigg|_{E^{(0)}(t)}\approx\frac{\partial S^{BC}}{\partial E|_\nu^{(0)}}. \quad (26)$$

This is an analog of the energy-balance equation - an equation of path balance $\left(\sum_{i=2}^{\nu}d\ln E|_{i-1}^{(0)}\right)/T_\int+dE|_\nu^{(0)}/T=dS^{BC}$. Returning from equation (19a) to equation (19) we similarly obtain $\left(\sum_{i=2}^{\nu}dS_{[E]}|_{i-1}^{(0)}\right)/T_\int+dE|_\nu^{(0)}/T=dS^{BC}$. This equation could be obtained directly by the differentiation of equation (14) as the logarithm of equation (9).

Above we utilized the assumption of ‘equiconnectivity’. We assumed that all microlevels of a given macrolevel have identical connectivities to other macrolevels. If this assumption is not true for a system, we would have to split energy macrolevels into sub-macrolevels gathered by the connectivity properties. Then, similarly to how we varied the paths in the time-energy space, we would also have to vary the paths is the space of connectivity properties. Although this introduces major complications for the theory, it will provide the same narrow maximum of the probability of macropaths.

## 7. Random walk of excitations

In this section we will illustrate all concepts developed above for a simple example of random walk of excitations. As a model we consider a lattice of discrete sites (atoms). The total number of sites on the lattice is $N$ which is infinite in the thermodynamic limit. Linear size of the lattice equals $L \propto N^{1/d}$ sites and the boundary between the lattice and the thermobath consists of $N^{d-1/d}$ sites where $d$ is the dimensionality of embedding space.

We assume that each site (atom) has a discrete set of energy levels, all separated by the constant interval of energy $\Delta E$: energy levels of an atom are $0, \Delta E, 2\Delta E, 3\Delta E, \ldots$ Then the system has a spectrum of macrolevels $[0],[\Delta E],[2\Delta E],[3\Delta E],\ldots$ Macrolevel $[0]$ consists of only one microlevel $\{0\}$ where all atoms are on the $0^{st}$ energy levels. Macrolevel $[\Delta E]$ consists of $N$ microlevels $\{\Delta E\}$ where all atoms are on the $0^{st}$ energy levels except one which is on the $1^{st}$ energy level. Macrolevel $[2\Delta E]$ consists of $N!/((N-2)!2!)$ microlevels $\{2\Delta E\}$ where all atoms are on the $0^{st}$ energy levels except two which are on the $1^{st}$ energy level and $N$ microlevels $\{2\Delta E\}$ where all atoms are on the $0^{st}$ energy levels except one which is on the $2^{nd}$ energy level. And so on. For a macrolevel $[n\Delta E]$ the degeneracy of this macrolevel is $(N-1+n)!/((N-1)!n!)$. Following Boltzmann's approximation of a classical gas, we can assume that the number of excitations $n$ is large: $n >> 1$, infinite in the thermodynamic limit, has an order of but is much less than the number of degrees of freedom $N$: $n \propto N$, $n << N$. This allows us to assume that the occupation number of each level of atom's spectrum is much less than unity. Following this approximation we can neglect quantum effects of multiple identical excitations of the same site and approximate the degeneracy of the macrolevel $[n\Delta E]$ as

$$g_{[n\Delta E]} = \frac{(N-1+n)!}{(N-1)!n!} \approx_{\ln} \frac{N!}{(N-n)!n!} \approx_{\ln} \frac{1}{(1-n/N)^{N-n}(n/N)^{n}} \tag{27}$$

We limit connectivity of our model so that excitations can move only between adjacent pairs of sites. Interaction with the thermobath is limited to appearance and disappearance of excitations at the boundary sites. When the system moves from a microlevel $\{n\Delta E\}$ to a microlevel $\{n'\Delta E\}$, new *n'* excitations of the new microlevel are organized by all possible moves of previous *n* excitations in *q* possible directions to *q* adjacent sites, where *q* is the coordination number of the lattice. Also $k$ excitations at the boundary sites could go to the thermobath while $(n'-n)+k$ excitations should come from the thermobath to the boundary sites. Therefore the number of possible paths leading from the microlevel $\{n\Delta E\}$ to the macrolevel $[n'\Delta E]$ with logarithmic accuracy is $q^{n'}$ possible moves of the remaining excitations times $(N^{(d-1)/d}+k)!/(N^{(d-1)/d}!k!)$ as a choice of leaving excitations times $(N^{(d-1)/d}+(n'-n)+k)!/(N^{(d-1)/d}!((n'-n)+k)!)$ as a choice of coming excitations. We assume that the system is in a state close to the equilibrium state with the thermobath and we could expect that on the average less than one excitation can leave or come at each boundary site. Therefore both the second and the third multipliers have an order of $N^{(d-1)/d}!$ which we with logarithmic accuracy neglect in comparison with the first multiplier $q^{n'}$: $N^{(d-1)/d} << n \propto N$. This gives us the number of paths, leading from the microlevel $\{n\Delta E\}$ to the macrolevel $[n'\Delta E]$, as

$$g_{\{n\Delta E\}\to[n'\Delta E]} \approx_{\ln} q^{n'} \equiv g_{\{\}\to[n'\Delta E]} \,. \tag{28}$$

We should notice a peculiar fact that the number of paths, leading from a microlevel, does not depend on the energy of this microlevel of initiation. This happed because the possible change of energy is limited to $N^{(d-1)/d}\Delta E$ and in this close range the connectivity of the model does not depend on the choice of the initial microlevel.

The probability for a microlevel $\{n\Delta E\}$ to move to a microlevel $\{n'\Delta E\}$ is

$$w^{BC}_{\{n\Delta E\}\to\{n'\Delta E\}} \equiv \frac{1}{Z^{BC}(\{n\Delta E\})}\exp(-(n'-n)\Delta E/T). \tag{29}$$

where $Z^{BC}(\{n\Delta E\}) \equiv \sum_{\{n\Delta E\}\to\{n'\Delta E\}:\{n\Delta E\}\ is\ given}\exp(-(n'-n)\Delta E/T)$ is the partition function for the initiation from the given microlevel $\{n\Delta E\}$. Recalling that our connectivity is limited to the energy range $N^{(d-1)/d}\Delta E$ where it is independent of the destination point, we obtain

$$Z^{BC}(\{n\Delta E\}) \approx_{\ln} \sum_{(n'-n)=0,\pm 1,\ldots,\pm N^{(1-d)/d},\ n\ is\ given} g_{\{n\Delta E\}\to[]}\exp(-(n'-n)\Delta E/T). \tag{30}$$

where $g_{\{n\Delta E\}\to[]} \approx_{\ln} q^n$ is the number of paths, leading from a given microlevel to any macrolevel in the possible energy range $N^{(d-1)/d}\Delta E$. We should definitely differentiate why we used $g_{\{\}\to[n'\Delta E]} \approx_{\ln} q^{n'}$ in (28) and $g_{\{n\Delta E\}\to[]} \approx_{\ln} q^n$ in (30). The situation is analogous to the case of the equilibrium canonical ensemble. The partition function depends on the initial state of exodus while the degeneracy, competing with the probability, is the degeneracy of the final state of destination. However, we always can assume that $g_{\{n\Delta E\}\to[]} \approx_{\ln} q^n \approx_{\ln} g_{\{\}\to[n\Delta E]}$.

The number of terms in sum (30) has an order of $N^{(d-1)/d}$, each of them has an order of $g_{\{n\Delta E\}\to[]}\exp(-N^{(1-d)/d}\Delta E/T)$, therefore with the logarithmic accuracy we obtain

$$Z^{BC}(\{n\Delta E\}) \approx_{\ln} g_{\{n\Delta E\}\to[]} \equiv q^n \,. \tag{31}$$

If we consider a macropath $[n|_0 \Delta E]\to[n|_\nu \Delta E]$, the probability of each micropath $\{n|_0 \Delta E\}\to\{n|_\nu \Delta E\}$ is

$$w^{BC}_{\{n|_0 \Delta E\}\to\{n|_\nu \Delta E\}} = \frac{1}{g_{[n|_0 \Delta E]}} \prod_{i=1}^{\nu} \frac{1}{g_{\{n|_{i-1}\}\to[]}} \exp\left(-(n|_i - n|_{i-1})\Delta E/T\right) \approx_{\ln}$$

$$\approx_{\ln} \frac{1}{g_{[n|_0 \Delta E]}} \exp\left\{-\sum_{i=1}^{\nu} n|_{i-1}\, \Delta E \frac{\ln q}{\Delta E} - (n|_\nu - n|_0)\Delta E/T\right\} \equiv \tag{32}$$

$$\equiv \frac{1}{Z^{BC}} \exp\left\{-\sum_{i=1}^{\nu} n|_{i-1}\, \Delta E/T_\int - n|_\nu\, \Delta E/T\right\},$$

where $T_\int \equiv \Delta E/\ln q$. So, for the case of random walk of excitations we obtain Boltzmann distribution of final energies and of integrals of evolutions of energies. This model gives Feynman's path integral formalism exactly.

For the degeneracy of the macropath $[n|_0 \Delta E]\to[n|_\nu \Delta E]$ we obtain

$$g_{[n|_0 \Delta E]\to[n|_\nu \Delta E]} \approx_{\ln} g_{[n|_0 \Delta E]} \prod_{i=1}^{\nu} g_{\{\}\to[n|_i \Delta E]} \approx_{\ln} g_{[n|_0 \Delta E]} \exp\left\{\sum_{i=1}^{\nu} n|_i \ln q\right\}. \tag{33}$$

The competence of (32) and (33) for the probability

$$W^{BC}_{[n|_0 \Delta E]\to[n|_\nu \Delta E]}[E(t)] = g_{[n|_0 \Delta E]\to[n|_\nu \Delta E]} w^{BC}_{\{n|_0 \Delta E\}\to\{n|_\nu \Delta E\}}[E(t)] \propto g_{\{\}\to[n|_\nu \Delta E]} \exp\left(-n|_\nu\, \Delta E/T\right) \tag{34}$$

to observe the macropath $[n|_0 \Delta E]\to[n|_\nu \Delta E]$ in the ensemble gives a very narrow maximum and a very narrow minimum for the path free energy action: $\Phi^A_{[n|_0 \Delta E]\to[n|_\nu \Delta E]} = -\ln\left(Z^{BC} W^{BC}_{[n|_0 \Delta E]\to[n|_\nu \Delta E]}\right)$. We see that in the case of our model the probability

of a macropath is equivalent to the probability of its final state, as if in the equilibrium canonical ensemble, only instead of the degeneracy of that state we should use its incoming connectivity.

## 8. Conclusions

In this paper we have developed the formalism of non-equilibrium statistical mechanics for non-equilibrium phenomena. Far from the state of equilibrium we switched from the states to the paths to base our theory on the most basic quantities which directly determine the probability ensembles. We developed non-equilibrium statistical mechanics for the path ensembles and found the equation of equilibrium path, the path balance equation, the expression for the path entropy and the path free energy potential. Also we showed that an ensemble of systems can be described in terms of the effective temperatures. Although we used the non-equilibrium canonical ensemble to illustrate all concepts developed, our results have general applicability to any other ensemble. Another important result of this paper is that we generalized Gibbs principle of the minimization of the free energy potential for path ensembles, only in this case instead of characteristics of the states we had to move to characteristics of the paths.

## Appendix A

Let us consider as an example the case of equilibrium Gibbs statistical mechanics and, in particular, an equilibrium grand canonical ensemble at constant temperature $T = \text{const}$, constant pressure $P = \text{const}$, and constant chemical potential $\mu = \text{const}$ as BCs. Then we have three 'effective' temperatures: $T$, $T/P$, and $-T/\mu$ in the equilibrium distribution of probabilities

$$w^{BC}_{\{E,V,N\}} = \frac{1}{Z^{BC}} \exp\left( -\frac{E}{T} - \frac{V}{T/P} - \frac{N}{(-T/\mu)} \right) \tag{A1}$$

where $\{E,V,N\}$ refers to a microstate with energy $E$, volume $V$, and $N$ particles. We choose one of the temperatures, $T$, to be explicitly present ahead of the exponent in (A1) to represent the free energy:

$$\Omega \equiv T\left( \frac{E}{T} + \frac{V}{T/P} + \frac{N}{(-T/\mu)} - S \right) = E + PV - \mu N - TS \, . \tag{A2}$$

As $dE = TdS - PdV + \mu dN$ for quasi-equilibrium processes, for the differential of (A2) we obtain the habitual equation

$$d\Omega = VdP - Nd\mu - SdT \, . \tag{A3}$$

Although this equation provides relations for the derivatives of the free energy, we still can obtain its more symmetric form. We define the 'action' of the free energy as the minus exponent of the probability distribution (A1) minus the entropy

$$\Phi^{\Omega} \equiv \frac{\Omega}{T} \equiv \frac{E}{T} + \frac{V}{T/P} + \frac{N}{(-T/\mu)} - S \, . \tag{A4}$$

For its differential we obtain

$$d\Phi^{\Omega} = Ed\left(\frac{1}{T}\right) + Vd\left(\frac{1}{T/P}\right) + Nd\left(\frac{1}{-T/\mu}\right). \tag{A5}$$

This form is clearly more symmetric because all extensive, variable quantities are accompanied by the differentials of the respective constant temperatures as it is supposed to be for the differential of the free energy action. Sometimes we cannot prefer one temperature among others. In particular, this situation occurs in statistical mechanics of multifractals [23], where the distribution of probabilities is a symmetric form of all temperatures.

Therefore the action of the free energy is more suitable to be employed instead of the free energy itself. The difference is a constant multiplier which does not influence the minimization principle.

Of course, the grand canonical ensemble was used here only as an example, and the discussion above works for any other ensemble.

**Appendix B**

In Gibbs equilibrium statistical mechanics an equilibrium state is found as a minimum of a free energy potential. For the equilibrium microcanonical ensemble the free energy potential is the negative entropy

$$-S \equiv <\ln w_{\{\}}> \equiv \sum_{\{\}} w_{\{\}} \ln w_{\{\}} \,. \qquad \text{(B1)}$$

In other words, to find the equilibrium distribution of probabilities we need to minimize (B1) with the constraint $\sum_{\{\}} w_{\{\}} = 1$. This is called the principle of entropy maximization.

For the equilibrium canonical ensemble the same principle of entropy maximization is also often used [24]. In this case to find the equilibrium distribution of probabilities we need to minimize (B1) with two constraints $\sum_{\{\}} w_{\{\}} = 1$ and $\sum_{\{\}} w_{\{\}} E_{\{\}} = E^{BC}$ where $E^{BC}$ is the value of energy prescribed by the thermobath. This minimization gives the correct Boltzmann distribution, however we could find this approach to be artificial. Indeed, why should we apply the second constraint $\sum_{\{\}} w_{\{\}} E_{\{\}} = E^{BC}$ when nothing restricts the system to consume as much energy as it would like and only on average to correspond to the thermobath? Applying constraints, we would expect them to act on the level of microstates but not on the level of averaged quantities.

The reason is that the true free energy potential that should be minimized for the canonical ensemble is the Helmholtz energy

$$A \equiv < E_{\{\}} > + T < \ln w_{\{\}} > \equiv \sum_{\{\}} w_{\{\}} [E_{\{\}} + T \ln w_{\{\}}] \qquad \text{(B2)}$$

or its action

$$\Phi^{A} \equiv A / T \equiv < E_{\{\}} > / T + < \ln w_{\{\}} > \equiv \sum_{\{\}} w_{\{\}} [E_{\{\}} / T + \ln w_{\{\}}]) \qquad \text{(B3)}$$

but not the entropy [10]. Indeed, in this case we obtain the correct Boltzmann distribution using only the 'true' constraint $\sum_{\{\}} w_{\{\}} = 1$.

Although both methods give the same result, the minimization of the free energy potential (or its action) seems to be more beautiful scientifically because the free energy potential (or its action) directly refers to the distribution of probabilities as a minus logarithm of this distribution. Also this approach appears to be more fundamental because it uses the true boundary constraint of constant temperature, applicable on the level of microstates, and not the secondary quantities like the averaged energy of the ensemble.

## References

[1] Boltzmann L, *Weitere studien über das Wärmegleichgewicht unter Gasmolekulen*, 1872 *Wien. Ber.* **66** 275

[2] Boltzmann L, *Bemerkungen über einige Probleme der mechanischen Wärmetheorie*, 1877 *Wien. Ber.* **75** 62

[3] Boltzmann L, *Über die Beziehung zwischen dem zweiten Hauptsatzes der mechanischen Wärmetheorie und der Wahrscheinlichkeitsrechnung respektive den Sätzen über das Wärmegleichgewicht*, 1877 *Wien. Ber.* **76** 373

[4] Gibbs J W, *On the equilibrium of heterogeneous substances*, 1876 *Trans. Conn. Acad.* **3** 108

[5] Gibbs J W, *On the equilibrium of heterogeneous substances*, 1878 *Trans. Conn. Acad.* **3** 343

[6] Onsager L, *Reciprocal relations in irreversible processes. I*, 1931 *Phys. Rev.* **37** 405

[7] Onsager L, *Reciprocal relations in irreversible processes. II*, 1931 *Phys. Rev.* **38** 2265

[8] Grandy W T, 1988 *Foundations of Statistical Mechanics* vol II: Nonequilibrium Phenomena (Dordrecht: D. Reidel Publishing Company)

[9] Grandy W T, 2008 *Entropy and the Time Evolution of Macroscopic Systems* (Oxford: Oxford University Press)

[10] Abaimov S G, *Applicability and non-applicability of equilibrium statistical mechanics to non-thermal damage phenomena*, 2008 *J. Stat. Mech.* P09005

[11] Abaimov S G, *Applicability and non-applicability of equilibrium statistical mechanics to non-thermal damage phenomena: II. Spinodal behavior*, 2008 *J. Stat. Mech.* P03039

[12] Onsager L and Machlup S, *Fluctuations and irreversible processes*, 1953 *Phys. Rev.* **91** 1505

[13] Kikuchi R, *Irreversible cooperative phenomena*, 1960 *Ann. Phys.* **10** 127

[14] Kikuchi R, *Variational derivation of the steady state*, 1961 *Phys. Rev.* **124** 1682

[15] Lavenda B H and Cardella C, *On the persistence of irreversible processes under the influence of random thermal fluctuations*, 1986 *J. Phys. A: Math. Gen.* **19** 395

[16] Dewar R, *Information theory explanation of the fluctuation theorem, maximum entropy production and self-organized criticality in non-equilibrium stationary states*, 2003 *J. Phys. A: Math. Gen.* **36** 631

[17] Woo H-J, *Statistics of nonequilibrium trajectories and pattern selection*, 2003 *Europhys. Lett.* **64** 627

[18] Evans R M L, *Driven steady states: rules for transition rates*, 2004 *Physica A* **340** 364

[19] Evans R M L, *Rules for transition rates in nonequilibrium steady states*, 2004 *Phys. Rev. Lett.* **92** 150601:1

[20] Evans R M L, *Detailed balance has a counterpart in non-equilibrium steady states*, 2005 *J. Phys. A: Math. Gen.* **38** 293

[21] Kikuchi R, *Path integral in irreversible statistical dynamics*, 1961 *Phys. Rev.* **124** 1691

[22] Abaimov S G, *Non-equilibrium statistical mechanics of non-equilibrium damage phenomena*, 2009 *arXiv* **0906.0190**

[23] Abaimov S G, *Application of classical statistical mechanics to multifractals and dynamical systems*, 2008 *arXiv* **0805.0347**

[24] Jaynes E T, *Information theory and statistical mechanics*, 1957 *Phys. Rev.* **106** 620

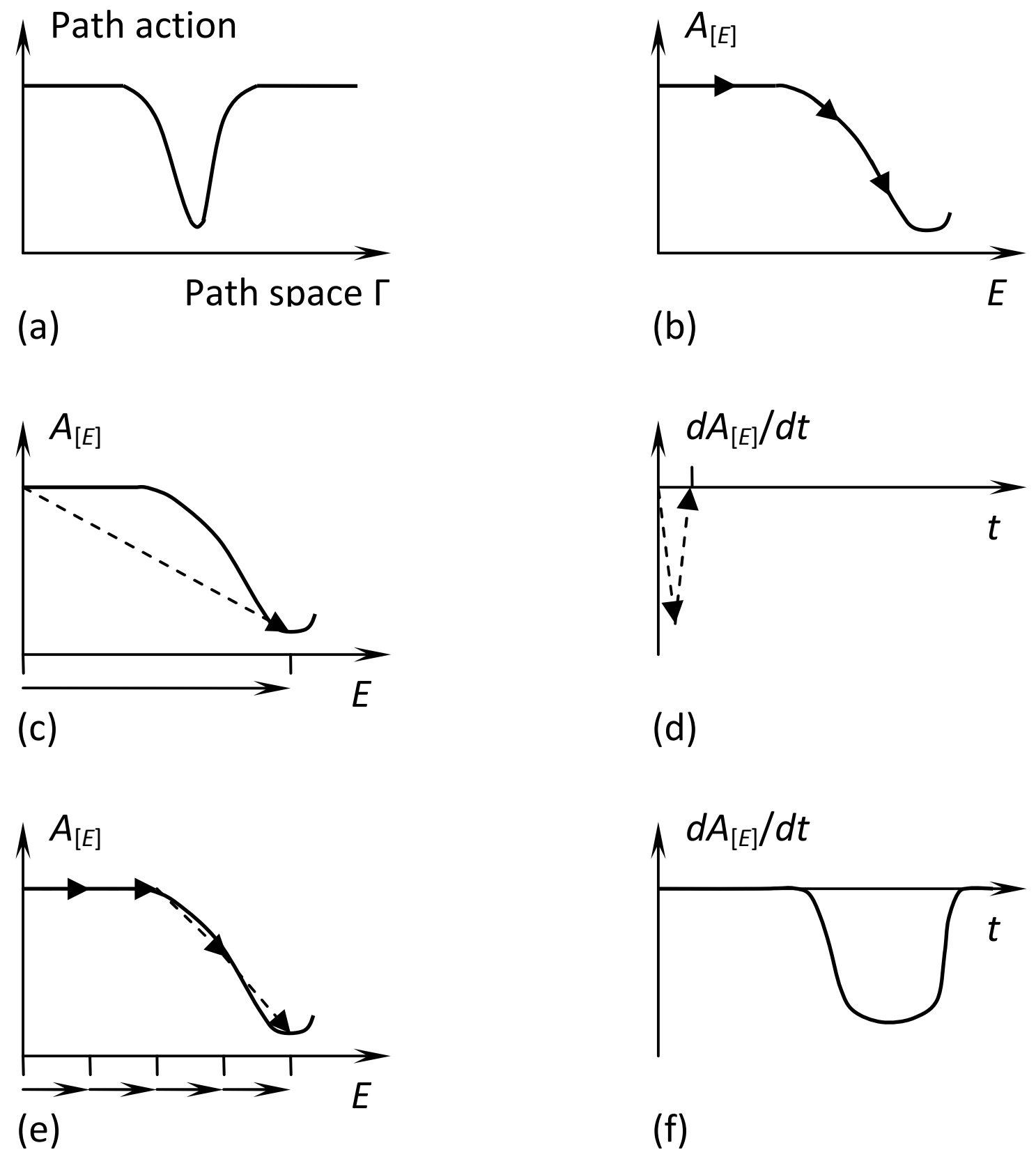

Figure 1. A schematic illustration of (a) minimum of a path action functional over the space of all possible paths; (b) usual tendency of a system to decrease its free energy along the equilibrium path; (c-d) a system with unlimited connectivity; (e-f) a system with limited connectivity.